# On Yields of May 11, 1998 Indian explosions by network averaged teleseismic P-wave spectra

S.K.Sikka

Office of the Principal Scientific Adviser to Government of India, Delhi

Abstract

We show here that the network averaged teleseismic P-wave spectra for Indian explosions of May 11, 1998, given by Barker et al, do not have an unambiguous interpretation. Barker et al had earlier demonstrated these were similar to the Shagan River testing site of former Soviet Union. We prove here that these are equally consistent with RUBIS (57 kt) and PILEDRIVER (62kt) explosions in French Hogger and US Nevada testing sites respectively.

Three underground nuclear explosions were detonated by India on 11 May, 1998. These were triggered simultaneously and comprised of a thermonuclear device, a fission device and a subkiloton device in spatially separated shafts. Initial estimates of the yields of about 60 kt were derived from close in measurements and from analysis of regional and teleseismic data. In 1998, Barker et al [1], in a paper in Science, applied the technique of network averaged teleseismic P-wave spectra from recordings from some stations to show that the observed shape of the May 11, 1998 Indian explosions (5/11/98, often referred to as POK-2 explosions) is remarkably similar to that for the tests at the former Soviet Semipalatinsk site and completely inconsistent with US Nevada test site The amplitude versus frequency data plotted by them were communicated by Science to Sikka, Roy and Nair, in response to their comments to the paper of Barker et al ( see Fig.1).

This analysis led Barker et al to apply the $m_b$ versus yield relation for explosions at Shagan site

$$m_b = 4.45 + 0.75 \log Y \qquad (1)$$

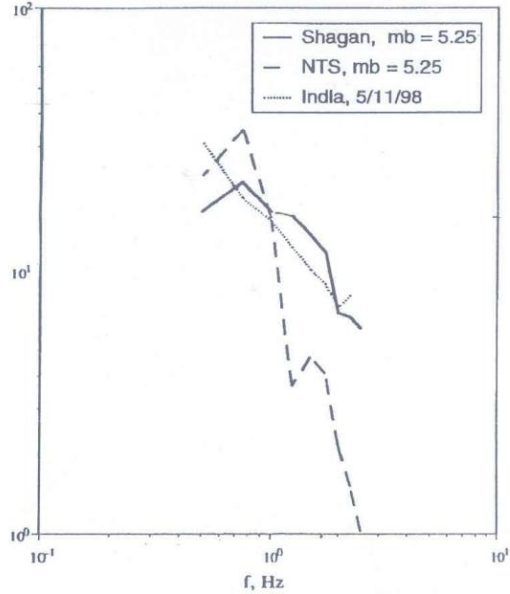

Fig.1: Networked averaged P wave spectrum for POK-2 explosions compared to the test at Semipalatinsk test site and the Nevada test site as given by Barker et al [1].

to the POK-2 data and gave the yield for POK 2 as only 12 kt. This was much less than the value of about 60 kt announced by Indian scientists (Sikka and Kakodkar [2]). This resulted into a furor in scientific and political circles around the globe and a conclusion that the thermonuclear device tested underperformed.  Later on, BARC scientists carried out the radio-active logging of the test sites and firmly established that the fusion reactions had indeed taken place as exemplified by the peaks for $^{54}$Mn, $^{46}$Sc and $^{22}$Na corresponding to activation products of 14 MeV neutrons in measured γ- ray spectra on rock samples recovered at the thermonuclear emplacement site.  The plots of the ratio of the activation and fission products, e.g. $Mn^{54}/Ce^{144}$ etc., between the thermonuclear and fission tests are very telling. Same is true for the $Eu^{152}/Eu^{154}$ ratio. These match with the computed neutron spectra assuming the observed yields.  From a   detailed analysis, Manohar et al [3] determined the yields for the thermonuclear and fission explosions as 50±10 kt and 13±3 kt respectively. Both the seismic and radioactive results about POK-2 have been summarized recently by Chidambaram [4]

. We show in this note that the data on POK-2 given by Barker et al in Fig.1 does not have a unique interpretation. It can be fitted to another test site, consistent with the yield of about 60 kt. It may be stated here that the rock surrounding the thermonuclear test was pink granite. We have carried out measurements of the constitutive properties of rock samples around the emplacement points of POK 2. In particular, the Hugoniot data of granite have been measured at TBRL by Manjeet Singh et al [5]. Comparison with other granite data from other regions demonstrates that the data of Pokhran granite is very similar to that of granite at the French Hogger testing site in Sahara.

This observation prompted us to re-examine the net work averaged P wave spectra of POK-2 as already shown in Fig.1 against data for an explosion of yield near 60 kt in Hogger granite. Spectra have been published by Murphy and Barker [6] for the 57 kt RUBIS explosion of 20/10/63 in Hogger granite. This was done by simply replotting the Pokhran data provided by Barker et al on the graph for this explosion (i.e. on Fig.12 of Murphy and Barker [6]). It also contains the data for the 62 kt US PILEDRIVER explosion in granite at Nevada (see fig. 2). Murphy and Barker have, in the same paper, proved that the granite source coupling and upper mantle attenuation in Nevada is close to French granite over this short-period band. .

The figure demonstrates that the attenuation amplitude with frequency of POK- 2 is also similar to RUBIS and PILEDRIVER and, thus, leads to a direct conclusion that the yield of POK-2 could be near to the yields of the said two explosions in granite i.e. near 60 kt and instead of 12 kt as put out by Barker et al [1]. This also confirms as already conclusively shown earlier by Sikka, Roy and Basu [7] that the relation (1) was not applicable to the Indian test site. This was illustrated in the $m_b$ versus $M_s$ plot of data of Shagan and Nevada explosions along with the values for Pokhran explosions of 1974 and 1998 (see Fig.1 of Sikka et al (7)).. A site bias of 0.4 $m_b$ units was indicated between the Shagan and Pokhran sites. This is in agreement with the $m_b$ versus yield formula for Hogger and Nevada granites

$$m_b = 3.93 + 0.89 \log Y \qquad (2)$$

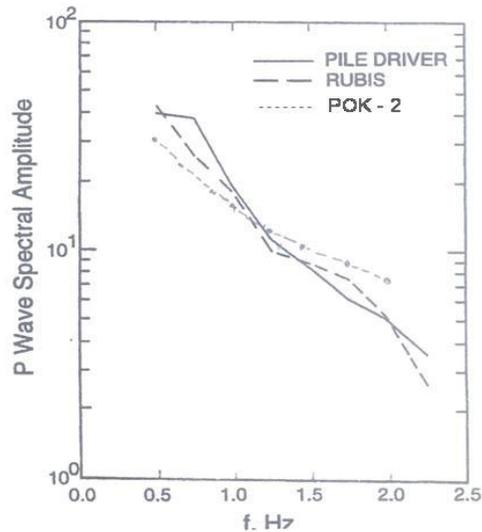

**Fig.2: The network averaged spectra of POK-2 explosions, as given in Fig.1 plotted against those for French Sahara RUBIS (57kt) and NTS PILEDRIVER (62kt) explosions. The data for the later two explosions is from Murphy and Barker [6].**

.

The above analysis clearly shows that the net work-averaged P wave spectra given by the US scientists now fully support the radiochemical yield estimates of Indian scientists and also the interpretation of close-in data and seismic analysis done earlier (Sikka et al [7- 9] and Roy et al [10]) and that the thermonuclear device performed as designed (see Table 1).

I thank my erstwhile colleagues in Seismology Division of Bhabha Atomic Research Centre for useful discussions.

1. Barker, B., Clark, M., Davis, P., Fisk, M.,Hedlin, M., Israelsson, H., Khalturin, V., Kim, W. Young, McLaughlin, K., Meade, C., Murphy, J., North, R., Orcutt, J., Powell, C., Richards, P., Stead, R., Stevens, J., Vernon, F. and Wallace, T. (1998), Science, Vol. 281, pp.1967.
2. Sikka, S.K. and Kakodkar, A. (1998 May) BARC Newsletter, No. 172.
3. Manohar, S.B., Toamr, B.S., S.S., Shukla, V.K., Kulkarni, V.V., Kakodkar, A. (1999 July), BARC Newsletter, No. 186.
4. Chidambaram,R. (2008), Atoms for Peace: An International Journal, Vol. 2, pp 41-58.
5. Manjeet Singh et al – unpublished
6. Murphy, J. and Barker, B. (2001), Pure Appl. Geophys. Vol.158, pp2123-2171.

7. Sikka, S.K., Roy, F. and Basu, T.K. (2002), Current Science, Vol. 83, No. 8, pp.992-997.
8. Sikka, S.K., Roy, F. and Nair, G.J. (1998b), Current Science, Vol. 75, No. 5, pp.486-491.
9. Sikka, S.K., Nair, G.J., Roy, F., Kakodkar, A. and Chidambaram, R. (2000), Current Science, Vol. 79, No. 9, pp.1359-1366.
10. Roy, F., Nair, G.J., Basu, T.K., Sikka, S.K., Kakodkar, A., Bhattacharya, S.N. and Ramamurthy, V.S. (1999), Current Science, Vol. 77, No. 12, pp.1669-1673.

Table1. Yields of POK-2 nuclear explosions determined by various techniques

| Technique | Yield (kt) |
|---|---|
| Close-in-acceleration | 58 |
| P-wave magnitude calibrated with Pok-1 yield | 53 |
| Net- work averaged P wave spectrum (reinterpreted) | 57-62 |
| $\Delta m(Lg)$ between Pok 1 and Pok 2 | 58 |
| Ms from regional and teleseismic stations | 46-53 |
| $\Delta m$ (pcp) between Pok 1 and Pok 2 | 58-63 |
| Radiochemical     Thermonuclear     Fission | 50 ± 10 <br> 13 ± 3 |